\documentclass[10pt,twocolumn,twoside]{osajnl}

\usepackage{caption}
\DeclareCaptionFormat{myformat}{#1#2#3\hrulefill}
\usepackage{physics}

\journal{jocn} 

\setboolean{shortarticle}{false}

\title{Multiple reflection assisted Laser Doppler Vibrometer setup for high resolution displacement measurement}

\author[1,$\dag$,$\ast$]{Nikhil Kumar Pachisia}
\author[1,$\ddag$,$\ast$]{Kajal Tiwari}
\author[1,$\S$]{Ritwick Das}

\affil[1]{School of Physical Sciences, National Institute of Science Education and Research, HBNI, Jatni-752050, India}
\affil[$\ast$]{N. K. Pachisia and K. Tiwari contributed equally to this work}
\affil[$\dag$]{nikhil.pachisia@niser.ac.in}
\affil[$\ddag$]{kajal.tiwari@niser.ac.in}
\affil[$\S$]{ritwick.das@niser.ac.in}

\dates{\today}

\begin{abstract}
Non-intrusive displacement measurement technique has a distinct edge over displacement sensors working in contact mode. Laser Doppler Vibrometry based schemes (LDV) offer such an advantage. However, the measurements are limited up to $\lambda/4$, where $\lambda$ is the wavelength of optical source. An experimental design of high resolution non-intrusive displacement sensor working on the principle of basic LDV is being presented here. Resolution enhancement has been achieved through phase multiplication using multiple reflections within a high-Q cavity created by a vibrating surface and a high reflection mirror kept parallel to each other. Displacement of 72 $nm$ has been measured with an error range of $\pm 14$ $nm$ through direct counting of characteristic peaks in ``half cycle'' of the interferogram. The experimental design offers a straightforward route to measure displacement and the measurement resolution is mainly limited by the reflectivity and physical dimension of moving source.

\end{abstract}

\setboolean{displaycopyright}{false}

\begin{document}

\maketitle

\section{INTRODUCTION}

Vibration measurements essentially comprise frequency, instantaneous velocity, acceleration and displacement measurements, which provide crucial information about the robustness of the design, resilience to vibrations and structural defects of the object under evaluation \cite{fu2001modal}. However, most of the schemes developed so far for modal parameter estimations, have one basic disadvantage; they primarily work in contact-mode which could be unfeasible and possibly, erroneous in a few situations. Mass-loading introduces errors and may cause substantial damage especially when testing delicate or small structures or highly damped non-linear materials \cite{castellini2006laser}. Additionally, component positioning and orientation restricts contact methods. Non-intrusive interferometric techniques for vibration measurements offers an attractive route to overcome the bottlenecks imposed by contact measurements. Laser interferometric techniques utilizing Doppler effect for measurement of fluid velocity was first reported by Yeh and Cummins \cite{yeh1964localized} at Columbia University in 1964. Since then, crucial developments have taken place in the field of laser interferometry-based devices and several sensing configurations have been proposed which are essentially derived from the principle of Laser Doppler effect. The impact of a `vibrating' object on laser interferometry is employed for analyzing its durability and quality. This technique is widely deployed testing of foot-over bridges and elevated pathways \cite{nassif2005comparison,tabatabai1998bridge}, monitoring the health of dynamic machine parts such as automotive components \cite{castellini2002vibration}, turbo machines \cite{oberholster2009online}, damage detection and bio-medical applications such as study of vibration in tympanic membranes, sensing mechanical cardiovascular activity, diagnostic tool for patients with abdominal aortic aneurysms \cite{goode1996laser,chen2010laser,schuurman2013feasibility}.\\

Laser Doppler Vibrometer (LDV) has been extensively used for small displacement measurement over a decade \cite{rothberg2017international}. As the name suggests, LDV utilizes Doppler-shift which occurs due to scattering of light from moving surface. It is a single point ‘axial’ vibrometer as it measures only the component of the velocity along the laser line-of-sight. LDV technique allows the data acquisition at different points by straightforward adjustments of optical parts and thereby, provides additional degrees of freedom to enhance the measurement resolution. In addition, it facilitates remote measurements which may not be physically accessible. LDV based systems extend measurement capabilities with respect to the classical transducers (such as accelerometers \cite{ferrero2016exploiting}, strain gauges, inductive, capacitive electrical sensors, Linear Variable Displacement transformer \cite{meydan1992linear}) through accurate measurement of modal parameters. Such characteristics are brought about by remote, non-intrusive, high-spatial resolution measurements with reduced testing time and improved performance. There are quite a few modified versions (inbuilt designs by Polytec Inc., USA) of LDV are now available, which are equipped with a video camera and a scanning and tracking system. The scanning laser Doppler vibrometer (SLDV) \cite{sriram1990scanning} is an instrument based on laser interferometry that can measure high spatial resolution vibration data by sequentially positioning the measurement laser beam at discrete positions using two scanning mirrors (for the vertical and horizontal deflection of the beam) \cite{sels20173d}.

For nanoscale displacement sensing and estimation, a resolution better than $\lambda/2$ is required. A plausible alternative is improved version of a `self-mixing' laser having an integrated and portable system which could reliably attain the quantum detection regime (in `$nm$' scale) \cite{giuliani2002laser,scalise2002laser,koelink1992laser}. Sensing and measurement by using the laser relaxation oscillation (RO) frequency have been reported, which can lead to higher resolution and long range displacement sensing \cite{liu2017displacement}. However, the measurement range, in their technique, is limited by the distance between laser diode and vibrating object. In addition, a high laser diode injection current and hence a high laser power is crucial for improving resolution. An absolute distance measurement using optical combs derived from a femtosecond pulse laser have been proposed \cite{jin2007absolute}.

\begin{figure}
\centering
{
  \includegraphics[width=0.45\textwidth]{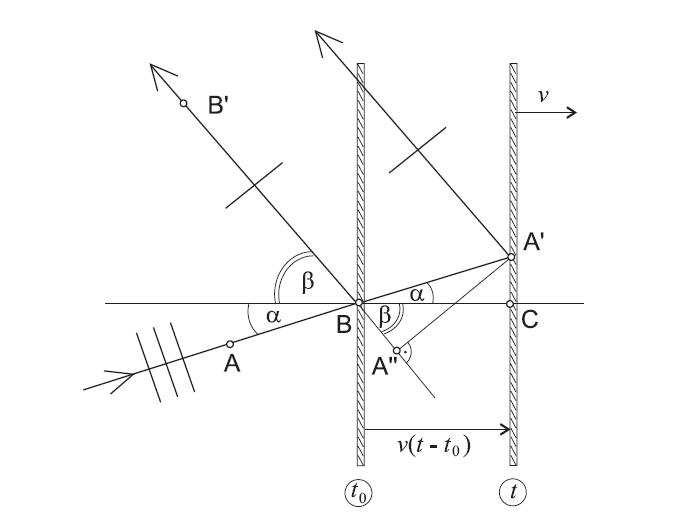}
}
 \begin{small}
\caption{A mirror moving at a constant velocity \textbf{v} towards right.}
\label{fig:1}
\end{small}
\end{figure}

In the present work, we present a simple, cost-effective experimental configuration of an LDV which is based on the principle of Laser-Doppler effect and Mach-Zehnder interferometry, and is capable of implementing non-contact measurements of frequency, single-point velocity and maximum displacement of vibrating objects. The present scheme elucidates the improvements in the basic design, methodology and strategy for measuring nanoscopic displacement range with high precision using phase multiplication of multiple reflections \cite{rembe2012there}. A combination of retro-reflector corner cube in conjunction with a plane mirror have been employed as reflecting objects for displacement and frequency measurements. A low power ($\leq~5~mW$) He-Ne laser which is generally used for interferometric purposes has been deployed as the source and vibrations are induced on the surface of reflectors through piezoelectric crystal.

\begin{figure}
\centering
{
  \includegraphics[width=0.4\textwidth]{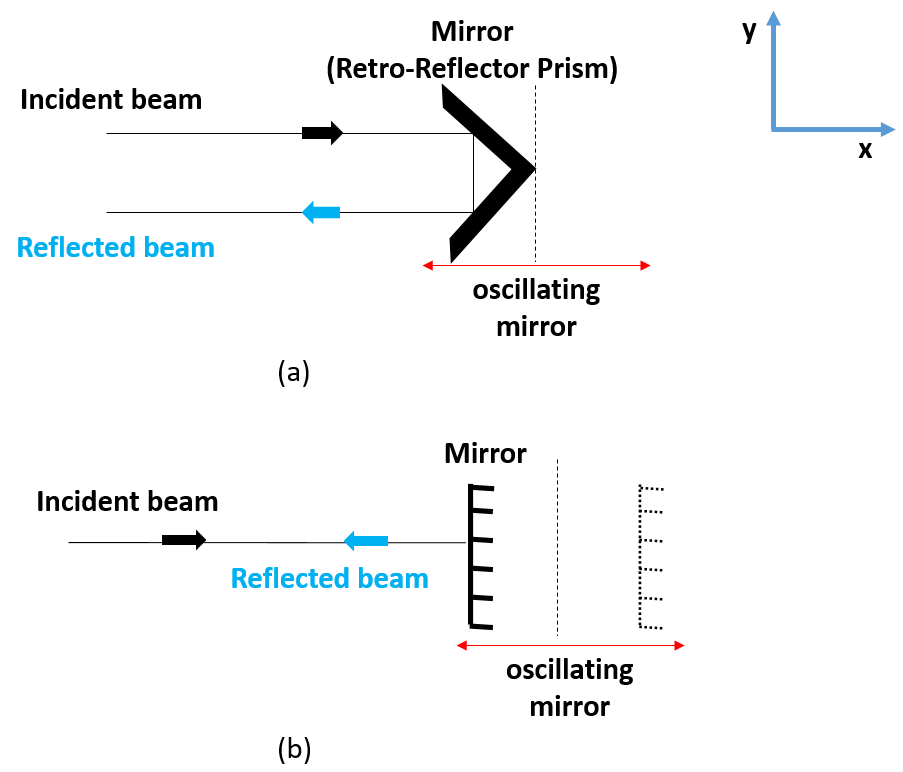}
}
 \begin{small}
\caption{(a) A retro-reflecting mirror mounted on a piezoelectric crystal undergoing oscillatory motion in the direction of propagating laser beam. (b) An equivalent configuration of ``(a)'', as angle between incident beam and reflected beam is $180^{o}$, implying $\alpha=0$ (refer Fig. \ref{fig:1}).}
\label{fig:2}
\end{small}
\end{figure}
\section{THEORY}
\subsection{Doppler effect due to moving mirror} \label{theory_a}
We note the frequency shift in a collimated He-Ne laser beam reflected from a mirror moving with uniform velocity via Doppler effect \cite{gjurchinovski2005doppler}. Consider the schematic shown in Fig. \ref{fig:1} where, the frequency of reflected beam is given by,
\begin{equation}
    f = f_{o}\qty(\frac{1 - 2 \frac{v}{c} \cos{\alpha } + \frac{v^{2}}{c^{2}}}{1-\frac{v^{2}}{c^{2}}})
    \label{eq:1}
\end{equation}
where,
$f_{o}$ is the frequency of incident laser beam, $v$ is the velocity of mirror, $c$ is the speed of light, $\alpha$ is the angle at which light is incident on the mirror from the direction of motion of the mirror (refer Fig. \ref{fig:1}).

We consider a case where object is moving towards the direction of propagation of laser beam. For any arbitrary direction (of movement/vibration), the component of displacement in the direction of the beam will be used for analysis due to the axial nature of LDV. Consider a mirror moving uniformly in positive $x$ direction (right), then the frequency ($f$) of reflected beam is given by Eq. (\ref{eq:1}) with $\alpha = 0$ (refer Fig. \ref{fig:2}) which is the case when mirror as well as the laser beam are moving in same direction.

\begin{figure}
\centering
{
  \includegraphics[width=0.45\textwidth]{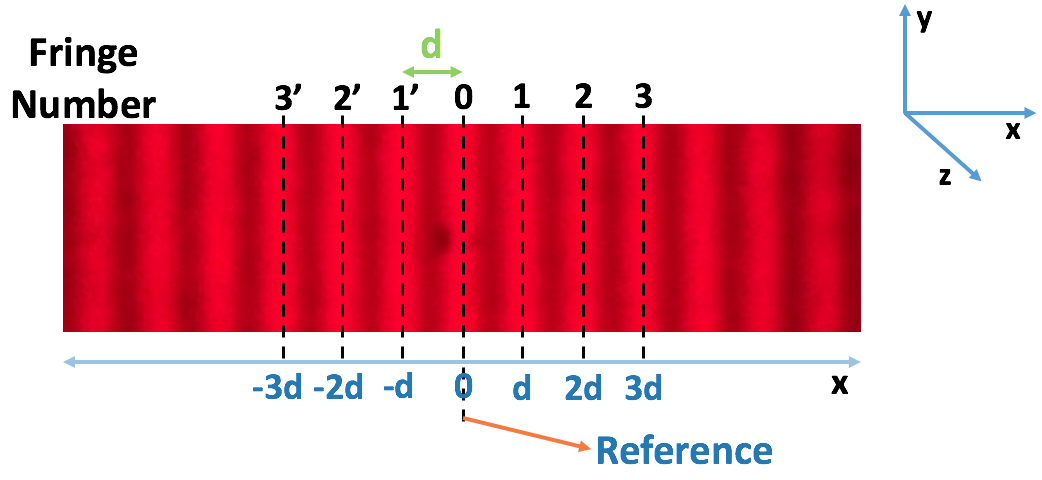}
}
 \begin{small}
\caption{Fringe pattern when the retro-reflecting mirror at rest position. Assuming center of bright fringe "$0$" as reference, we mark the distance of each fringe as a multiple of `$d$'}
\label{fig:3}
\end{small}
\end{figure}
\begin{figure}
\centering
{
  \includegraphics[width=0.48\textwidth]{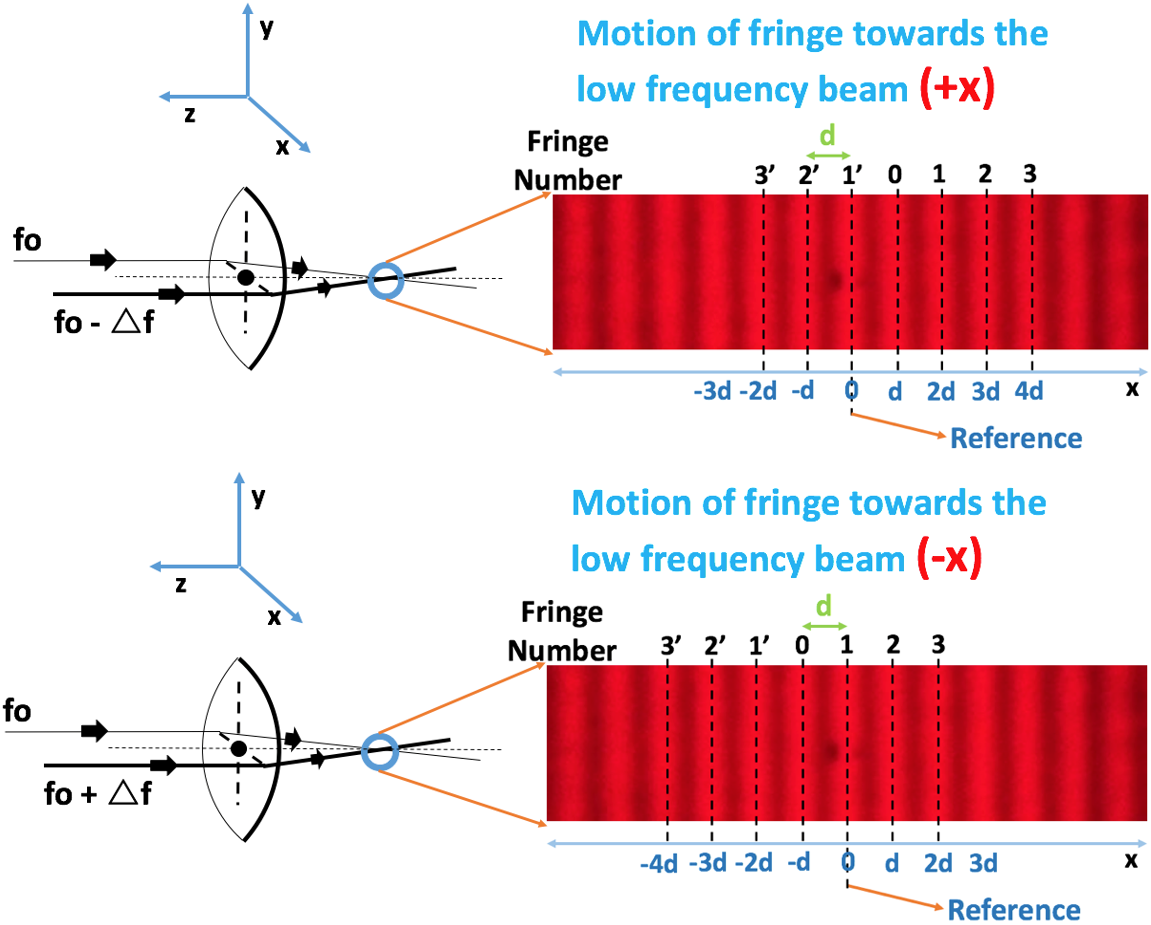}
}
 \begin{small}
\caption{Moving fringes created by two laser beams with slightly different frequencies (due to Doppler shift).}
\label{fig:4}
\end{small}
\end{figure}

Therefore, $f$ will be given as,
\begin{equation*}
    f = f_{o} \qty(\frac{1 - 2 \frac{v}{c} + \frac{v^{2}}{c^{2}}}{1-\frac{v^{2}}{c^{2}}})
\end{equation*}
\begin{equation*}
    f = f_{o}\qty(\frac{c-v}{c+v})
\end{equation*}
The shift in frequency can be written as: 
\begin{equation*}
    \Delta f = f_{o} - f = f_{o}-f_{o}\qty(\frac{c-v}{c+v})
\end{equation*}
since $c>>v$,
\begin{equation}
    \Delta f = \frac{2v f_{o}}{c} = \frac{2v}{\lambda}
    \label{eq:2}
\end{equation}

where wavelength $(\lambda$) is 632.8 $nm$ in case of He-Ne laser.
The reflected beam will have a frequency $(f_{o}-\Delta f)$. Similarly for the case when the velocity is towards left, the frequency of reflected beam will be $(f_{0} + \Delta f)$ where the value of $\Delta f$ is given by Eq. (\ref{eq:2}).

\begin{figure}
\centering
{
  \includegraphics[width=0.48\textwidth]{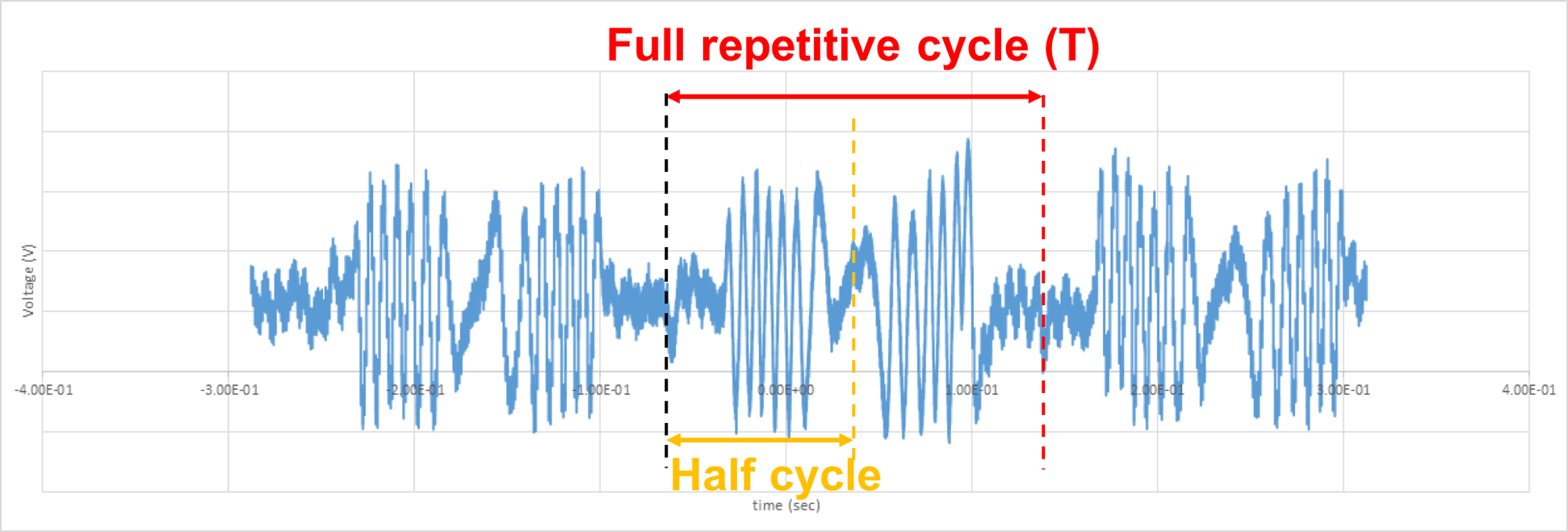}
}
 \begin{small}
\caption{A typical oscilloscope trace when retro-reflecting prism oscillates.}
\label{fig:5}
\end{small}

\end{figure}

\subsection{Moving fringes}
From Section (\ref{theory_a}), we observe that frequency of laser beam is shifted by $\pm \Delta f$ amount after reflection from a mirror moving in $-x$/$+x$ respectively. When the mirror is not vibrating, a static fringe pattern is observed, refer Fig. \ref{fig:3}.\\

It should be noted that the translation of one mirrors in a 'Michelson' interferometer results in movement of the interference fringes in one direction. In other words, the direction of motion could be ascertained through the direction of movement of fringes which primarily dictates the path difference. On a similar note, when the mirror begins to vibrate, the fringes were observed to follow a pattern conforming to the movement of mirror. Fringes move towards the lower frequency beam, with velocity ($v_{f}$) given by,
\begin{equation}
    v_{f} = d_{f} \times \Delta f
    \label{eq:3}
\end{equation}
where, $d_{f}$ is the fringe width between two consecutive light or dark fringe, $\Delta f$ is frequency shift, given by Eq. (\ref{eq:2}).

From the geometry, it is apparent that the reflected beam frequency will be given as \textbf{$f_{o} - \Delta f$} for the half cycle when the mirror is moving towards the right $i.e.$ in direction of incident beam and the reflected beam frequency will be \textbf{$f_{o} + \Delta f$} when the mirror is moving towards left (towards the incident beam), where $f_{o}$ is the emission frequency of incident laser beam.

From Fig. \ref{fig:4}, we observe that the fringe pattern carries the signature of the back and forth motion of the mirror. Here, it is worthwhile to recall that the velocity is maximum at the mean position and becomes zero at either boundaries. The frequency shift and velocity of the mirrors are related by the Eq. (\ref{eq:2}) where $\Delta f$ turns out to be a function of time. The movement of fringes are related to $\Delta f$ through Eq. (\ref{eq:3}) and consequently, the fringes follow an oscillatory motion i.e. movement ceases at the boundaries, while being maximum at the mean position.

\subsection{Determining the oscillation frequency}
\label{theory_c}
Consider a single oscillation cycle of the retro-reflector prism, which is also represented in the Fig. \ref{fig:5}. When the retro-reflector prism moves in the direction of incident beam, then `$n$' number of bright or dark fringes cross the photodetector plane. Similarly, when it moves in a direction opposite of incident laser beam, an identical `$n$' number of bright or dark fringes cross the photodetector plane. Therefore, the oscilloscope signal repeats twice in one ``full oscillation cycle''.\\

The ``full cycle'' mentioned in Fig. \ref{fig:5} is referred to the oscillation of mirror from one extreme end to the other end and back to same position (Refer Fig. \ref{fig:6}). The zero signal region corresponds to an instantaneous rest position. The frequency of vibration for the object will be same as the frequency of the ``full cycle'' observed in the oscilloscope, given by 
\begin{equation}
    f_{vib} = \frac{1}{T}
    \label{eq:4}
\end{equation}
where, T is Time period of a "Full Cycle".

\subsection{Determining the displacement} \label{theory_d}
\begin{figure}
\centering
{
  \includegraphics[width=0.5\textwidth]{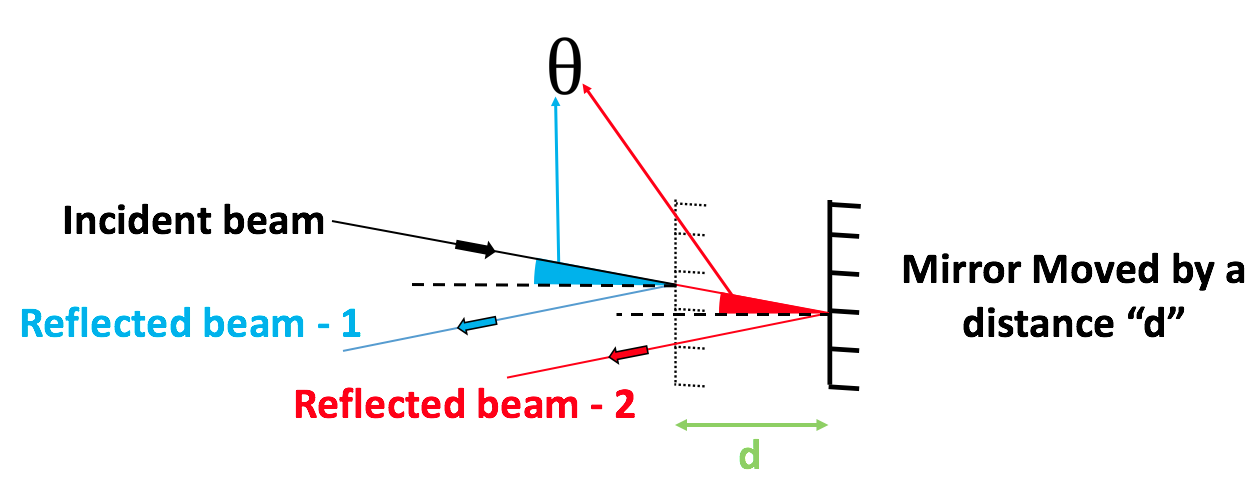}
}
 \begin{small}
\caption{Reflection of a laser beam by a mirror which is displaced by distance `$d$'.}
\label{fig:7}
\end{small}

\end{figure}

Consider a beam being reflected from a mirror which has been displaced by a distance `$d$' as shown in Fig. \ref{fig:7}. The path difference ($\Delta X$) between reflected beam-1 and reflected beam-2 is,
\begin{equation}
    \Delta X = \frac{2d}{\cos{\theta}}
    \label{eq:5}
\end{equation}

\begin{figure}
\centering
{
  \includegraphics[width=0.4\textwidth]{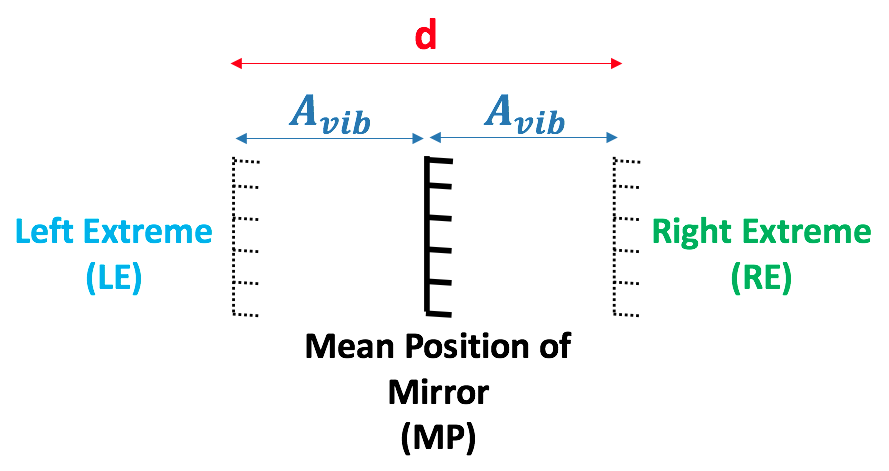}
}
 \begin{small}
\caption{A schematic representation of motion of an oscillating mirror}
\label{fig:6}
\end{small}

\end{figure}

where $\theta$ is the angle of incidence of beam on the mirror. Consider a `Michelson' interferometer set-up, with mirror adjusted in such a way that a bright/dark spot (constructive/destructive interference) is at the observation point. Now, if the mirror is moved such that the nearest bright/dark spot appears at the observation point, the path difference ($\Delta X_o$) is given by
\begin{equation}
     \Delta X_o=\frac{2d_o}{\cos{\theta}}= \lambda
     \label{eq:6}
\end{equation}
where $d_o$ is the distance moved by mirror.
Hence,
\begin{equation}
     d_o=\frac{\lambda\times \cos{\theta}}{2}
     \label{eq:7}
\end{equation}
If `$N$' number of bright/dark fringes cross the observation point then, distance moved by mirror ($d$) is given by,
\begin{equation}
     d =\frac{N\times \lambda }{2}\cos{\theta}
     \label{eq:8}
\end{equation}

For a ``normal incidence'' $\theta = 0^{o}$, the maximum displacement of the mirror is related to  number of bright/dark fringes crossing the detector by the relation,
\begin{equation*}
    \Rightarrow d = (2) \times A_{vib} = N \times \frac{\lambda}{2} 
\end{equation*}
\begin{equation}
    A_{vib} = N \times \frac{\lambda}{4}
    \label{eq:9}
\end{equation}
where $A_{vib}$ is the maximum displacement made by the oscillating mirror from its mean position to either end, refer Fig. \ref{fig:6}. The factor of ``$2$'' is arising due to the fact that we are measuring the `$d$' when mirror moved from one extreme end (LE) to other extreme end (RE), as shown in Fig. \ref{fig:6}. `$N$' is the number of dark/bright fringes crossing the detector in a ``half cycle'' (Number of maxima/minima in half cycle on oscilloscope signal).  Please note that the ``half cycle'' mentioned in Fig. \ref{fig:5} as well as in the present context, is referred to the motion of mirror from one extreme end (say LE) to other extreme (say RE) and $\lambda$ is wavelength of light used ($632.8~nm$).
\\
\\
\section{EXPERIMENTAL SET-UP}
\begin{figure*}
\centering
{
  \includegraphics[width=1\linewidth]{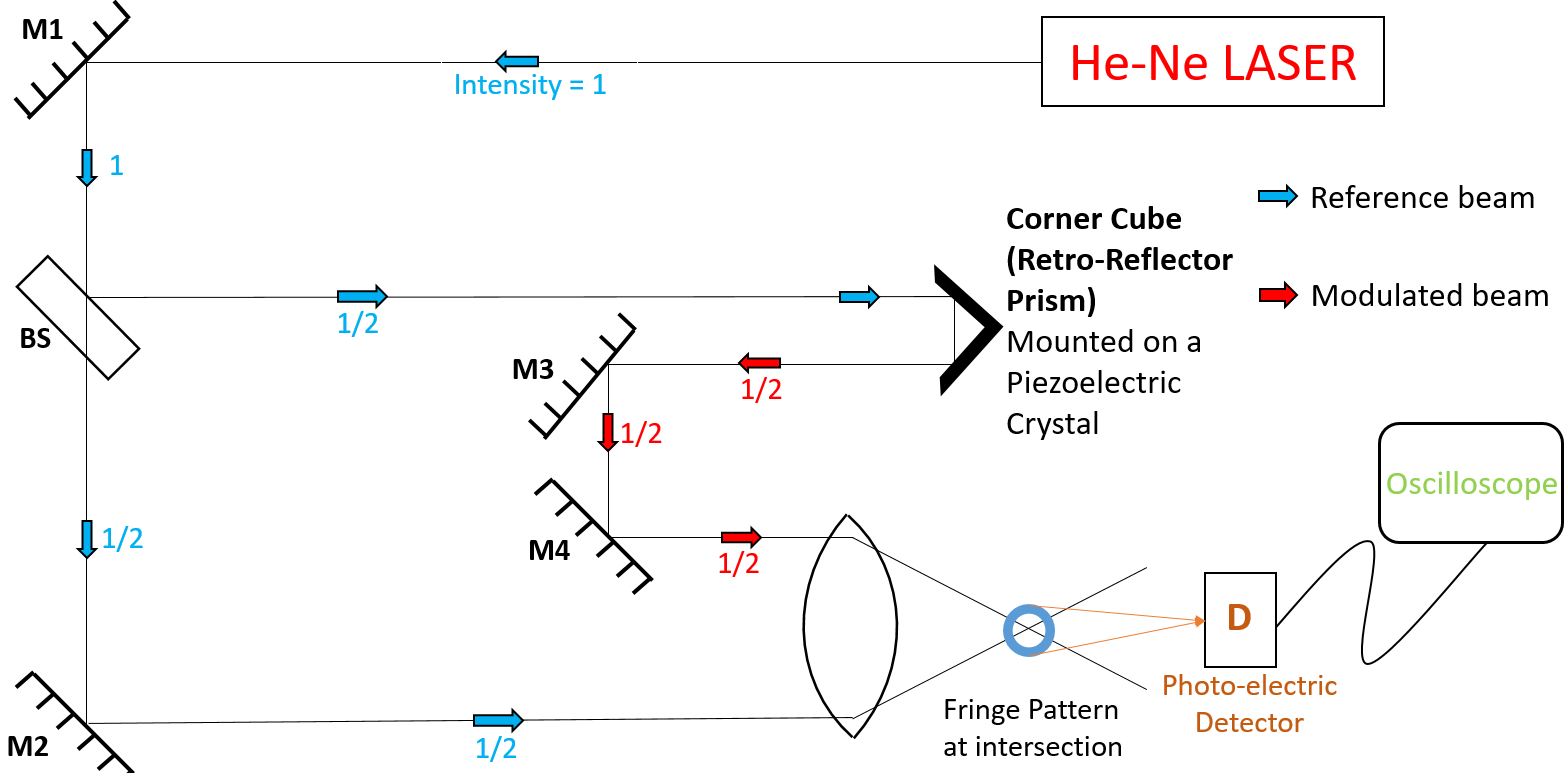}
}
 \begin{small}
\caption{A schematic diagram of the experimental set-up for measuring displacement and oscillation frequency. Please note the division of intensity in each arm.}
\label{fig:8}
\end{small}

\end{figure*}

A collimated beam from a He-Ne Laser (25-LHP-151-230, Melles Griot, Germany) beam is incident on a  non-polarizing beam splitter of splitting ratio (50:50). The deflected beam is incident on a retro-reflector (PS975M-A, Thorlabs, USA) prism (which reflects a incident beam by $180^{o}$) mounted on a piezo-electric crystal (Thorlabs, USA) with an actuator (TPZ001, Thorlabs, USA). The undeflected beam is taken to be the reference. The reflected and the reference beams are made to interfere after passing through a focusing lens as shown in Fig. \ref{fig:8}.

\begin{figure*}
\centering
{
  \includegraphics[width=1\linewidth]{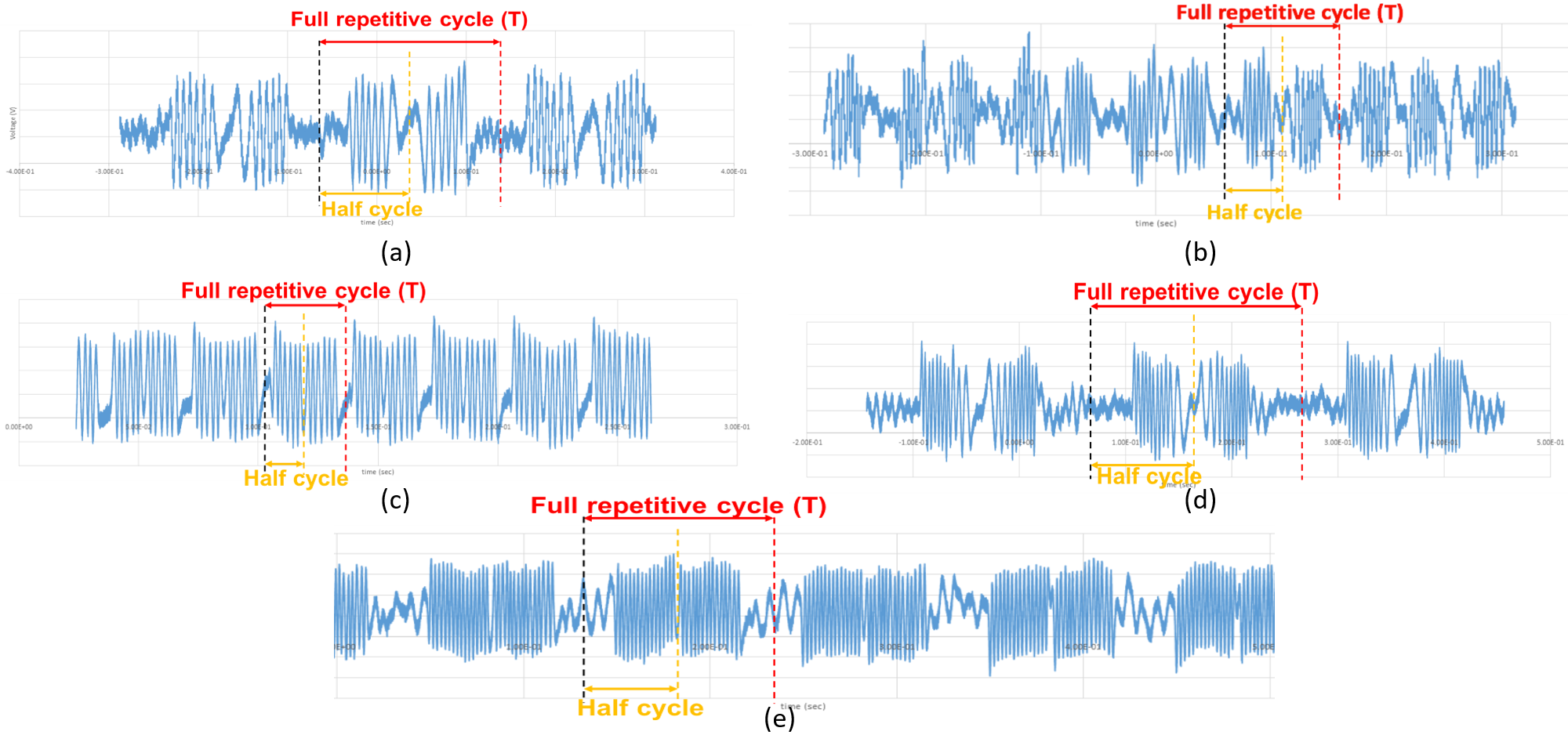}
}
 \begin{small}
\caption{Oscilloscope trace when the voltage applied to the piezo ($V_{pp}$) is fixed (a) $V_{pp}~=~5~V$, $f_{app}~=~5~Hz$. (b) $V_{pp}$ = 5V, $f_{app} = 10~Hz$. (c) $V_{pp} = 5~V$, $f_{app}~=~30~Hz$. When applied frequency to the piezo ($f_{app}$) is fixed, the oscilloscope trace (d) $V_{pp}~=~10~V$, $f_{app}~=~10~Hz$. (e) $V_{pp}~=~15~V$, $f_{app}~=~10Hz$}
\label{fig:9}
\end{small}

\end{figure*}

\begin{table*}
\centering
\begin{tabular}{p{2cm}p{2.1cm}p{2.1cm}p{2.3cm}p{2.3cm}p{2cm}p{2cm}}
\hline
\hline
Figure & Applied  & $f_{app}$ & Time & $f_{vib}$ ($ = \frac{1}{T}$)&  No. of peaks & $A_{vib}$ \\ 
No. & $V_{pp}$ & (Hz) & Period (T)  & (Hz) & in Half&$(N \times \frac{\lambda}{4})$ \\
 &  (Volt) & &(ms) & & cycle (N) & $\times 10^{3}$ nm \\ \hline
\multicolumn{7}{c}{Keeping $V_{pp}$ fixed} \\
\hline
i) & 5.0 & 5.0  & 200.0  & 5.0 & 6 & 0.95\\ 
ii) & 5.0 & 10.0  & 100.0  & 10.0 & 6 & 0.95\\ 
iii) & 5.0 & 30.0 & 33.3  & 30.0 & 6 & 0.95\\ \hline
\multicolumn{7}{c}{Keeping $f_{app}$ fixed} \\
\hline
ii) & 5.0  & 10.0 & 100.0 & 10.0 & 6 & 0.95\\ 
iv) & 10.0 & 10.0 & 100.0 & 10.0 & 10 & 1.55\\ 
v) & 15.0 & 10.0 & 100.0 & 10.0 & 15 & 2.37\\ \hline\hline
\end{tabular}
\caption{Table showing dependence of $N$ on $V_{pp}$ and $f_{app}$ }
\label{tab:1}
\end{table*}
The piezo-electric crystal is set to a desired frequency, waveform (sinusoidal) and amplitude through a waveform generator. Retro-reflector prism begins oscillating about its mean position leading to a to-and-fro motion of fringes. A fast photo-detector (UPD-200-SP, Alphalas, Germany) is positioned in the far-field (about 40 $cm$) from the point of intersection of the beams. This ensures a incidence of a fraction of the entire fringe pattern on the photo-detector window. A convex lens of an optimum focal length could also be employed for achieving the same. The detector output is monitored through an oscilloscope (RTB2004, Rohde \& Schwarz).
\begin{figure}
\centering
{
  \includegraphics[width=0.37\textwidth]{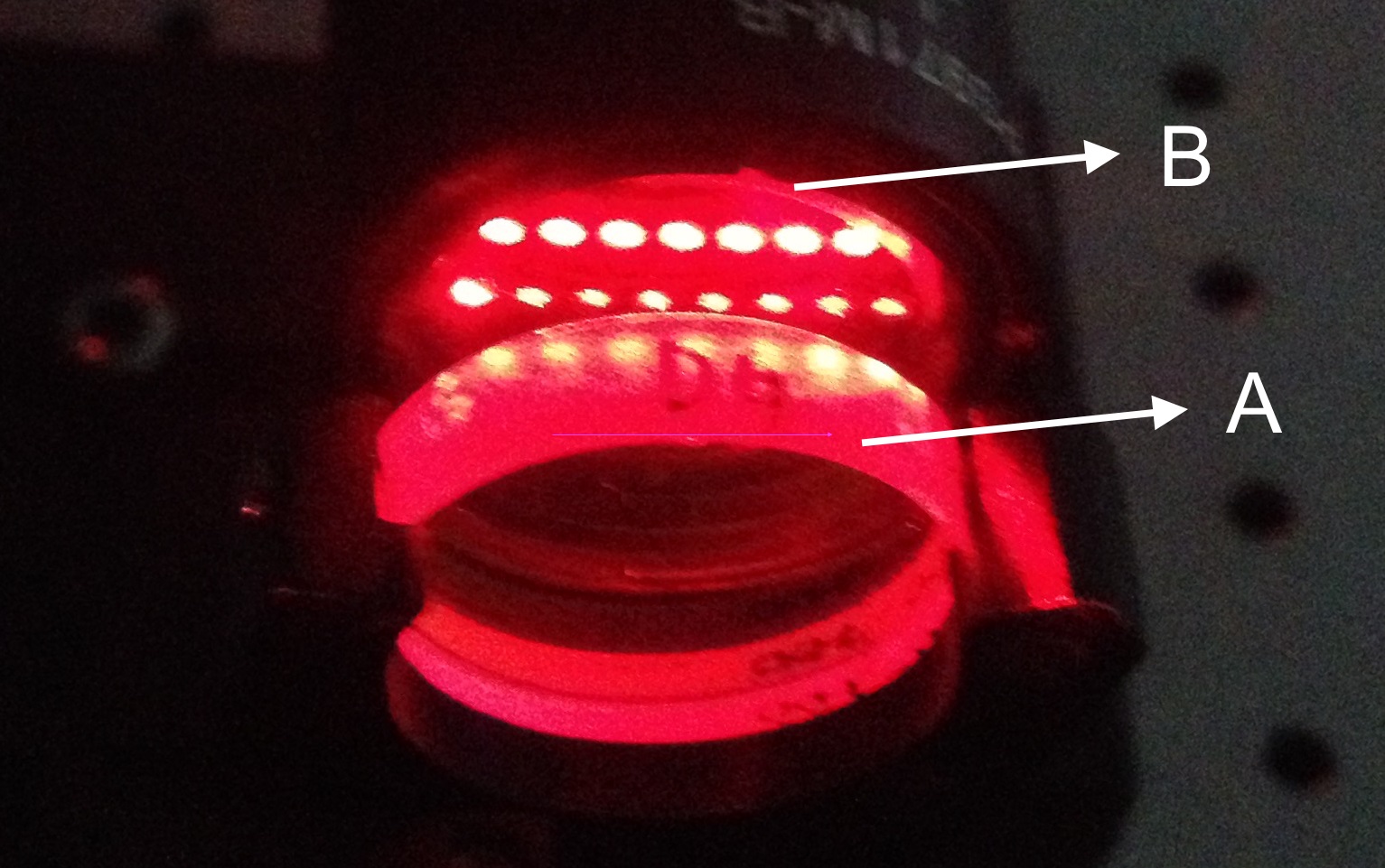}
}
 \begin{small}
\caption{Two mirrors A and B (with diameter $D~=~2.54~cm$) placed parallel and close to each other. Mirror B is mounted on a piezo crystal. The laser beam is reflected to mirror B at a small angle ($\theta$) for multiple reflections ($K~=~8$ in this figure).}
\label{fig:10}
\end{small}

\end{figure}

\section{OBSERVATIONS AND RESULTS}
We obtain the following signal on oscilloscope as shown in Fig. \ref{fig:9} by choosing different values of frequency ($f_{app}$) and amplitude of the vibration (peak-to-peak voltage $V_{pp}$) through the waveform generator which feeds signal to the piezo crystal. The vibration frequency of the prism has been calculated from the time period $T$ (recorded from oscilloscope) using the Eq. (\ref{eq:4}) and the amplitude of vibration ($A_{vib}$) for the prism is estimated using Eq. (\ref{eq:9}).

\begin{figure}
\centering
{
  \includegraphics[width=0.45\textwidth]{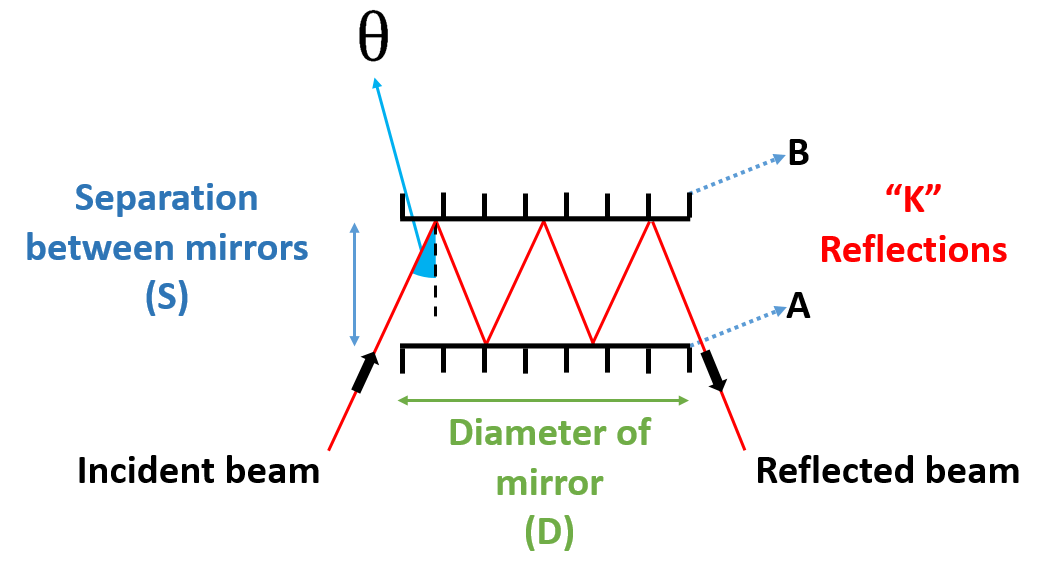}
}
 \begin{small}
\caption{Schematic diagram for computing approximate value of $\theta$ }
\label{fig:11}
\end{small}

\end{figure}

From Table-\ref{tab:1}, we observe that the calculated frequency ($f_{vib}$) is consistent with  ($f_{app}$). Also, for a fixed value of $V_{pp}$, the amplitude of the vibrating body does not change with change in driver frequency of oscillation. For $V_{pp}=0.8~V$, we observe that the number of peaks in a ``half cycle'' (N) is 1. Hence, $A_{vib} \approx 159~nm$ using Eq. (\ref{eq:9}).

\begin{figure*}
\centering
{
  \includegraphics[width=1\textwidth]{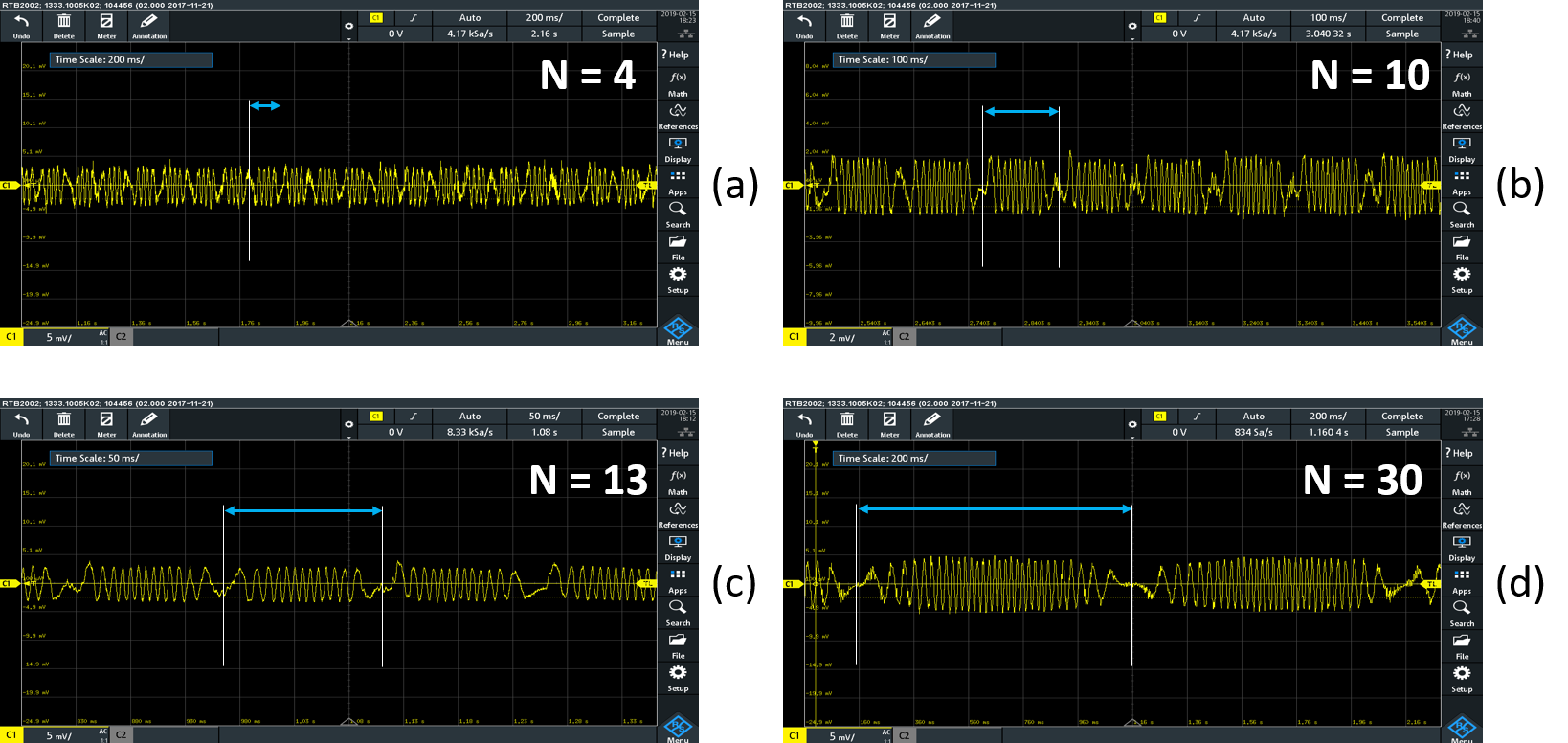}
}
 \begin{small}
\caption{Recorded oscilloscope traces for (a) $V_{pp}$ is \textbf{0.5 V} when $K = 9$, (b) $V_{pp}$ is \textbf{1 V} when $K = 10$, (c) $V_{pp}$ is \textbf{1.5 V} when $K = 8$, (d) $V_{pp}$ is \textbf{5 V} when $K = 5$}
\label{fig:12}
\end{small}

\end{figure*}
\section{ENHANCING THE RESOLUTION OF DISPLACEMENT MEASUREMENT}
It is apparent that the minimum $A_{vib}$ which we could measure using the aforementioned scheme, is $\lambda/4$, where amount of distance travelled during half cycle is $\lambda/2$. In order to enhance the resolution further, we need to increase the phase shift which can be done if multiple reflections in the retro-reflector is allowed by the experimental configuration. We achieve this by reflecting the laser beam multiple times from the mirror connected to the piezo-crystal which leads to accumulation of large phase shift in the exiting beam. Let us assume, we have `$K$' number of reflections from the vibrating mirror, refer Fig. \ref{fig:10}, where the displacement `$d$' moved by mirror (during oscillation) is very small as compared to the distance between the mirrors. `$K$' more or less remains unchanged when the mirror begins to vibrate. The path difference ($\Delta X$) between the beam reflected from the mirror when it is at an extreme end of oscillation cycle (say LE, refer Fig. \ref{fig:6}) and when it gets displaced by distance `$d$',

\begin{equation}
    \Delta X_K = K \times \frac{2d}{\cos{\theta}}
    \label{eq:10}
\end{equation}

Following an identical recipe discussed in Section (\ref{theory_d}), we obtain
\begin{equation}
    N \times \lambda= K \times \frac{2d}{\cos{\theta}}
    \label{eq:11}
\end{equation}
for ``$N$'' peaks in signal trace (or bright fringes).Also using Fig. \ref{fig:6} we get,
\begin{equation}
   (2) \times A_{vib} = d = \frac{N\lambda}{2\times K}\cos{\theta}
   \label{eq:12}
\end{equation}
Assuming the angle of incidence ($\theta$) to be small, a Taylor series expansion for $cos{\theta}$ gives us,
\begin{equation}
    A_{vib} = \frac{N\lambda}{4 \times K}\bigg(1- \frac{\theta^2}{2}+ O(\theta^4)\bigg)
    \label{eq:13}
\end{equation}
The zeroth-order amplitude correction would be given by,
\begin{equation}
    A_{vib}^{(0)} = \frac{N\lambda}{4 \times K}
    \label{eq:14}
\end{equation}
whereas the first-order correction term could be expressed as,
\begin{equation}
    A_{vib}^{(1)} = - \frac{N\lambda}{4 \times K}\times \frac{\theta^2}{2}
    \label{eq:15}
\end{equation}
\\
\begin{table*}
\centering
\begin{tabular}{*{16}{c}}
\hline
\hline
\multicolumn{1}{c}{} & \multicolumn{2}{c}{ 0.5 Volt } & \multicolumn{2}{c}{ 1 Volt }&  \multicolumn{2}{c}{ 1.5 Volt} & \multicolumn{2}{c}{ 2.0 Volt } &  \multicolumn{2}{c}{ 3.0 Volt } & \multicolumn{2}{c}{ 4.0 Volt } & \multicolumn{2}{c}{ 5.0 Volt } &\multicolumn{1}{c}{ Error Bar}\\ \hline
Reflections & N   & $A_{vib}$& N   & $A_{vib}$& N   & $A_{vib}$& N   & $A_{vib}$ &N   & $A_{vib}$ &N   & $A_{vib}$ &N   & $A_{vib}$  & $\Delta A_{vib} $ \\
(K) & & & & & & & & & & & & & & & $\qty(= \pm \frac{\lambda}{4\times K})$ \\ 
\hline
5         &     -- &    --         &  5    & 158  & 8  & 253      & 12 & 380      &  16 & 506   & 23 & 728     &  30 & 949 &  $\pm 32$    \\ 
6         &   --   &   --          & 6    & 158 & 9  & 237      & 14 & 369 
& 19 & 501 & 27 & 712      & 35 & 923  & $\pm 26$\\ 
7          & 3    & 68       & 7    & 158 & 11 & 249       & 16 & 362       & 23 & 520      & 34 & 768      & 45 & 1017  &  $\pm 23$        \\ 
8          &  --    &  --           & 8    & 158 & 13 & 257     & 19 & 376       & 26 & 514      & 40 & 791        & 51 & 1009  & $\pm 20$   \\ 
9          & 4    & 70 & 9    & 158 & 15 & 264 & 21 & 369 & 31 & 545 & 45 & 791 & 60 & 1055  & $\pm 18$\\ 
10         &   --   &  --           & 10   & 158 & 16 & 253      & 24 & 380      & 35 & 554      & 51 & 807     & 65 & 1028  &  $\pm 16$   \\ 
11         & 5    & 72 & 11   & 158 & 17 & 245 & 25 & 360 & 38 & 547 & 53 & 762 & 74 & 1064  &  $\pm 14$\\ \hline \hline
\end{tabular}
\caption{Values of $A_{vib}~(nm)$ recorded for increasing number of reflections at a fixed supplied $V_{pp}$.}
\label{tab:2}
\end{table*}
The angle of incidence $\theta$ could be measured directly from the experiment or alternately, noting the fact that $\theta$ depends on the diameter ($D$) of the mirror and the number of reflections ($K$) which could be directly counted (see Fig. \ref{fig:10}). In order to compute $\theta$, we propose a simple geometrical formulation (see Fig. \ref{fig:11}). The geometry shown in Fig. \ref{fig:11} results in,
\begin{equation}
    \tan{\theta}= \frac{D}{2 \times K \times S}
    \label{eq:16}
\end{equation}
where, $D$ is diameter of vibrating mirror, $S$ is separation between two mirrors, $K$ is Number of reflections by the vibrating mirror.\\

\begin{figure}
\centering
{
  \includegraphics[width=0.48\textwidth]{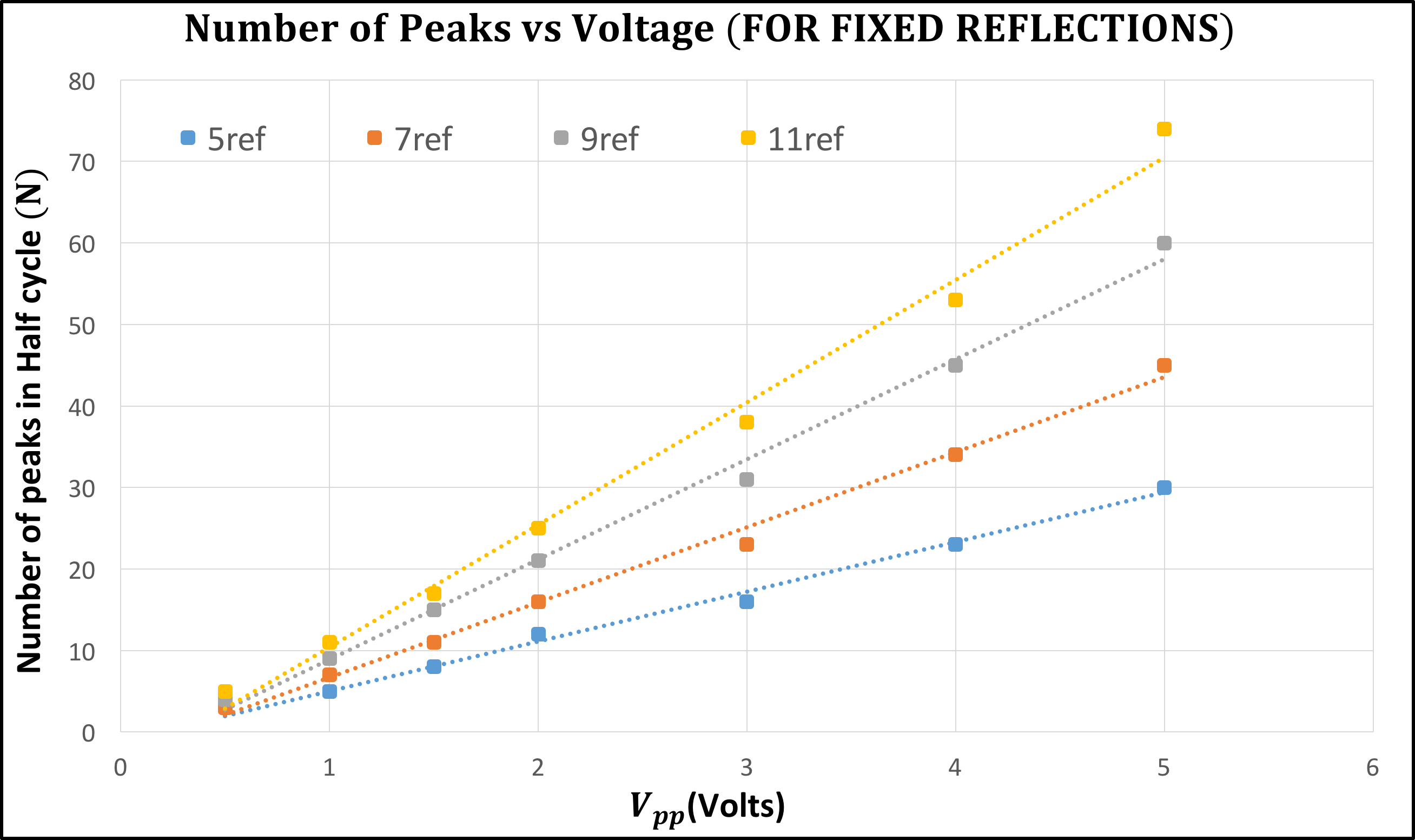}
}
 \begin{small}
\caption{Variation of $N$ as a function of applied voltage $V_{pp}$. Each coloured line represents a fixed number of reflections ($K$), refer Table-\ref{tab:2}. }
\label{fig:13}
\end{small}

\end{figure}

In this experiment, the value of $\theta < 10.0^{o}$ for all the observations. Due to this, the angular dependence of $A_{vib}$ could be truncated to first term only (see Eq. (\ref{eq:13})). In order to have a higher $K$ value, mirror ``A'' has been slightly tilted, which results in small change in $\theta$ for every reflection between mirrors ``A'' and ``B''. However, it has been ensured that $\theta$ does not exceed $10^o$ for any $K$. It is important to note that $A_{vib}$ is determined by the applied voltage on piezo-crystal and a higher voltage implies a greater displacement of the mirror $A_{vib}$. Fig. \ref{fig:13} confirms the linear dependence of ``number of peaks in half cycle'' ($N$) and the applied voltage ($V_{pp}$) on the piezo-crystal.

\begin{figure}
\centering
{
  \includegraphics[width=0.48\textwidth]{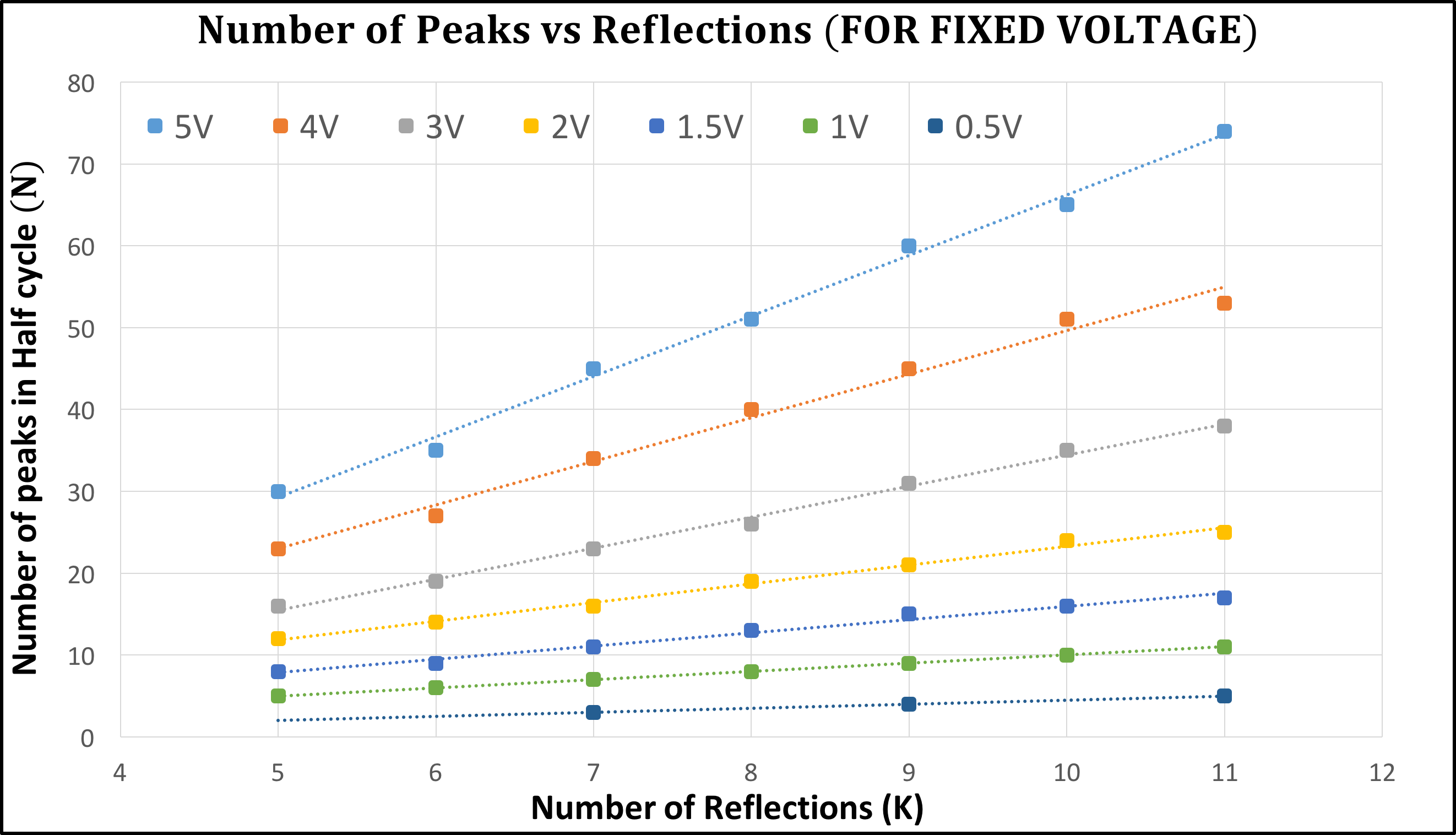}
}
 \begin{small}
\caption{Variation of $N$ as a function of $K$. Each coloured line represents a fixed peak to peak voltage $V_{pp}$, refer Table-\ref{tab:2}. }
\label{fig:14}
\end{small}

\end{figure}

Therefore, this allows us to obtain $A_{vib}$ (that depends on applied voltage $V_{pp}$) as a function of $N$ for a fixed $K$.\\

In Fig. \ref{fig:14}, we plot $N$ as a function of $K$ for a fixed $A_{vib}$ (that depends on Voltage applied $V_{pp}$). The variation indicates a linear dependence of $N$ on $K$ which is predicted by Eq. (\ref{eq:14}) as well.\\

\begin{figure}
\centering
{
  \includegraphics[width=0.48\textwidth]{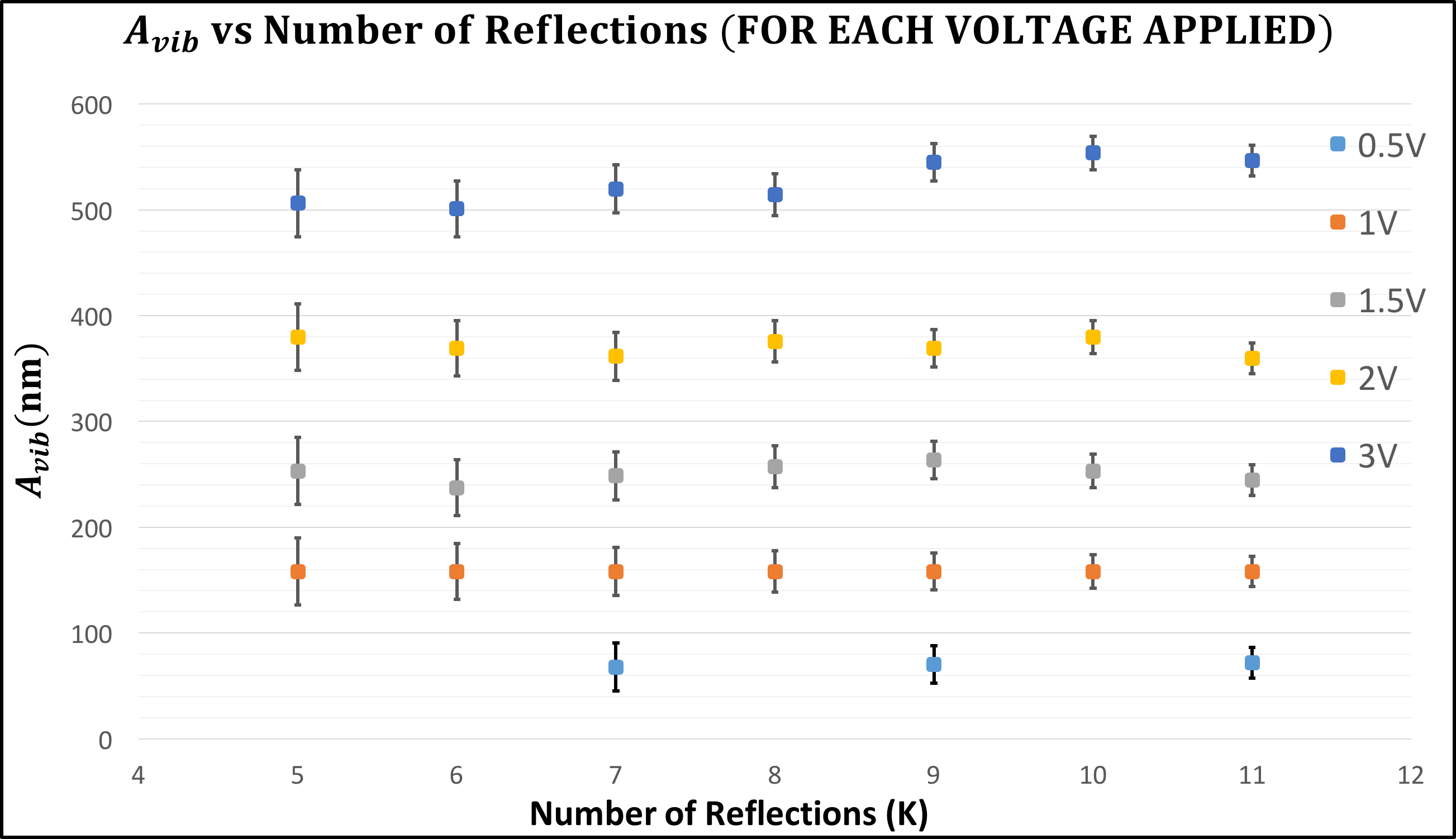}
}
 \begin{small}
\caption{Variation of $A_{vib}$ a function of number of reflections ($K$) with error bars for each $A_{vib}$,  refer Table-\ref{tab:2}. }
\label{fig:15}
\end{small}

\end{figure}

From Fig. \ref{fig:15}, the values of $A_{vib}$  are within error bounds. Error bar has been found pertaining to the fact that for a given $K$, smallest $\Delta A_{vib}$ is required to increase or decrease the value of $N$ (which is the only observable in our setup) by one unit is given by:
\begin{equation}
    \Delta A_{vib} = \pm \frac{\lambda}{4\times K}
    \label{eq:17}
\end{equation}
This defines the estimation error criteria for each measurement. Therefore, the actual value of displacement of the mirror ($A_{vib}$) will be within the error bar for a given voltage as shown in Fig. \ref{fig:15}. It is worth noting the fact that the error in each measurement is minimization through increasing $K$ (refer Eq. (\ref{eq:17})).\\

Figure \ref{fig:12} consist images of oscilloscope traces for a given $V_{pp}$ and $K$. ``$N$'' for Fig. \ref{fig:12}{\color{blue}(a)} is $4$ and the corresponding computed value of ``$A_{vib}$'' using Eq. (\ref{eq:14}) is $70 \pm 18~nm$. Following an identical route, ``$N$'' is 10 for Fig. \ref{fig:12}{\color{blue}(b)} and ``$A_{vib}$'' is estimated to be $158 \pm 16~nm$, ``$N$'' is 13 for Fig. \ref{fig:12}{\color{blue}(c)} which leads to $A_{vib}~=~257 \pm 20~nm$. For a given combination of $K$  and $N$, the values of $A_{vib}$ has been tabulated in Table-\ref{tab:2}. Using this multi-reflection scheme in an LDV, the minimum measurable distance would be $68\pm23~nm$ which could be improved further by re-appropriating the experimental configuration for obtaining smaller $N$. In fact, the distance measuring resolution is marginally affected due to this but the error in all the measurements reduces significantly. From Table-\ref{tab:2}, it could also be ascertained that $A_{vib}~=~158~nm$ for $V_{pp}~=~1~V$ for different value of $K$ which is essentially a consequence of the fact that the motion of mirror is retained in spite of altering the applied frequency $f_{supplied}$ for a given $V_{pp}$ applied to the piezo-driver. This is essentially a manifestation of the fact that $A_{vib}$ is only determined by the voltage applied to the piezoelectric crystal and remains unchanged for different values of $K$.

\section{CONCLUSION}
The multiple reflection scheme adopted in the LDV technique resulted in measurement of minimum displacement of $72~nm$ through direct counting of interference fringes. The minimum possible applied voltage $V_{pp}$ to the piezoelectric crystal through the driver (which was $0.5~V$ for our case), is the primary limiting factor which determines the minimum displacement that could be measured using this technique. The experimental design is straightforward and the linear relationship between the applied voltage (to vibrating/moving object) with the number of interference maxima/minima is  the key feature of the scheme. Linearity is maintained even in the sub-nanometer regime hence extrapolation is successfully possible. This technique is well suited for sinusoidal disturbances as well as superposition of multiple sinusoidal or any other waveform. It is worthwhile to note that the resolution as well as the rate of signal processing play a crucial role in determining the accuracy and sensitivity of the device. Further, it could be safely asserted that sub-nanometer sensitivity and resolution could be achieved using this technique through deployment of phase meters which are capable of resolving signals down to a few picometers \cite{kochert2012phase,hsu2010subpicometer}.


It is worth noting that the frequency applied to the piezo-crystal must be smaller for executing electromagnetic phase multiplication with large number of reflections. This is essentially due to the fact that the movement of fringes is rapid across the detector and consequently, a smaller time resolution is required to count the number of peaks in the signal trace. In other words, a better detector resolution and broader oscilloscope bandwidth would assist in resolving two closely-spaced maxima/minima of the interference fringes. Since the measurement is based on counting the number of peaks in oscilloscope trace, the error in every measurement is given by $\lambda/(4K)$. As discussed previously, this can be easily reduced several folds by increasing the number of reflections ($K$), which require a broader reflecting area for the mirror. An improvement in reflector design could enhance the resolution (and minimize the error in measurement) within the linear regime. The proposed configuration, due to its simple and cost-effective design, could deploy improved detection schemes and an appropriate signal processing technique for obtaining picometer-scale resolution. Simplicity and cost effectiveness of the setup is well suited for undergraduate laboratories.
\\

\section{ACKNOWLEDGEMENTS}
The authors extend their sincere gratitude to Anupa Kumari and Sushree Sahoo for providing the necessary equipment and technical assistance. They are grateful to the Department of Atomic Energy, Government of India for providing the necessary financial support.


\bibliographystyle{osajnl}

\end{document}